# Super-resolution multi-contrast unbiased eye atlases with deep probabilistic refinement

Ho Hin Lee,[a,†] Adam M. Saunders,[b,†,*] Michael E. Kim,[a] Samuel W. Remedios,[c] Lucas W. Remedios,[a] Yucheng Tang,[b] Qi Yang,[a] Xin Yu,[a] Shunxing Bao,[b] Chloe Cho,[d] Louise A. Mawn,[e] Tonia S. Rex,[e] Kevin L. Schey,[e,f] Blake E. Dewey,[g] Jeffrey M. Spraggins,[f,h] Jerry L. Prince,[g] Yuankai Huo,[a,b] Bennett A. Landman[a,b,d]

[a]Department of Computer Science, Vanderbilt University, Nashville, USA
[b]Department of Electrical and Computer Engineering, Vanderbilt University, Nashville, USA
[c]Department of Computer Science, Johns Hopkins University, Baltimore, USA
[d]Department of Biomedical Engineering, Vanderbilt University, Nashville, USA
[e]Department of Ophthalmology and Visual Sciences, Vanderbilt University Medical Center, Nashville, USA
[f]Department of Biochemistry, Vanderbilt University, Nashville, USA
[g]Department of Electrical and Computer Engineering, Johns Hopkins University, Baltimore, USA
[h]Department of Cell and Developmental Biology, Vanderbilt University, Nashville, USA

**Abstract**

**Purpose:** Eye morphology varies significantly across the population, especially for the orbit and optic nerve. These variations limit the feasibility and robustness of generalizing population-wise features of eye organs to an unbiased spatial reference.

**Approach:** To tackle these limitations, we propose a process for creating high-resolution unbiased eye atlases. First, to restore spatial details from scans with a low through-plane resolution compared to a high in-plane resolution, we apply a deep learning-based super-resolution algorithm. Then, we generate an initial unbiased reference with an iterative metric-based registration using a small portion of subject scans. We register the remaining scans to this template and refine the template using an unsupervised deep probabilistic approach that generates a more expansive deformation field to enhance the organ boundary alignment. We demonstrate this framework using magnetic resonance images across four different tissue contrasts, generating four atlases in separate spatial alignments.

**Results:** When refining the template with sufficient subjects, we find a significant improvement using the Wilcoxon signed-rank test in the average Dice score across four labeled regions compared to a standard registration framework consisting of rigid, affine, and deformable transformations. These results highlight the effective alignment of eye organs and boundaries using our proposed process.

**Conclusions:** By combining super-resolution preprocessing and deep probabilistic models, we address the challenge of generating an eye atlas to serve as a standardized reference across a largely variable population.

## 1 Introduction

Significant variation in human eye morphology, especially in the shape and size of the orbit and the optic nerve sheath diameter (ONSD), presents challenges in medical imaging to generalize population-wise features of eye organs to a spatial reference image. Different volumetric imaging modalities capture distinct perspectives on eye morphology. Typical imaging modalities include computed tomography (CT), magnetic resonance imaging (MRI), ultrasonography, and optical coherence tomography (OCT). The diversity of imaging protocols increases the amount of contextual information available. For example, researchers have used OCT to create a reproducible measure of the curvature of the eye.[1] Contrast agents injected into the vascular system can highlight abnormal tissues like lesions and tumors. In MRI, different imaging sequences result in different relaxation weightings, producing distinct tissue contrasts.

Even in healthy individuals, there is significant variation in orbit and optic nerve morphology. Differences in eye morphology have been associated with demographic variables like sex and ethnicity.[2] Researchers have used CT scans to find associations between orbital skull landmarks and sex and ethnicity.[3,4] A study examining ONSD in 585 healthy adults using ultrasonography found that the ONSD ranged from 3.30 mm to 5.20 mm and eyeball transverse diameter (ETD) ranged from 20.90 mm to 25.70 mm.[5] Similarly, another study with 300 healthy participants found that the ONSD diameter ranged from 5.17 mm ± 1.34 mm to 3.55 mm ± 0.82 mm at different locations in the intra-orbital space using CT imaging.[6] In addition, variation in eye morphology, particularly in the globe, depends on conditions that affect visual acuity, such as myopia and hyperopia. Researchers have used MRI to associate myopia with posterior eye shape.[2] A study examining differences in eye shape on MRI in emmetropia and myopia found that the globe is larger in all dimensions (with the largest changes axially followed by vertically then horizontally) as myopic refractive correction increases. Specifically, in myopia, the globe dimensions ranged from 22.1 mm to 27.3 mm axially, 21.1 mm to 25.9 mm vertically, and 20.8 mm to 26.1 mm horizontally. Even the typical emmetropic eye contains substantial variation across a population.[7,8]

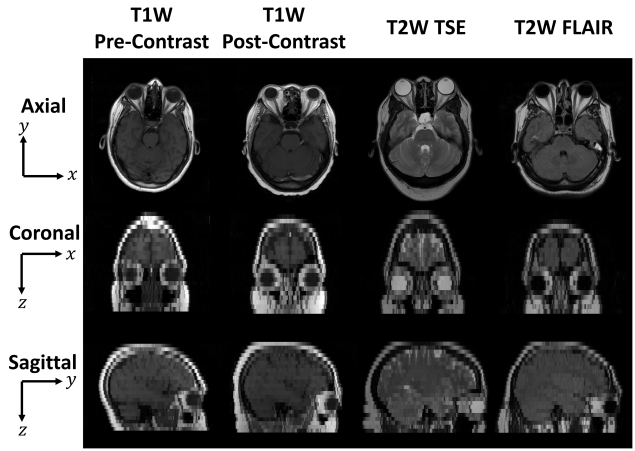

**Fig. 1** Representative in-plane (axial, first row) and through-plane (coronal, second row and sagittal, third row) slices for four MRI tissue contrasts from four different subjects. The coronal and sagittal through-plane slices are lower resolution than the axial in-plane slices and are visualized with nearest neighbor interpolation. The relatively lower resolution limits our ability to distinguish organs and generalize anatomical characteristics across populations.

The morphology of the eye is also import for understanding pathologies. Tumors like optical nerve sheath meningioma can compress the optic nerve, while optic nerve glioma can expand the optic nerve.[2] Thyroid eye disease can result in optical rectus muscle enlargement.[3] Changes such as these can be quantified using morphological metrics, e.g. the ONSD, which can be measured after segmenting the optic nerve from the surrounding orbital fat. These variations highlight the difficulty in creating a standardized reference image that is not biased by known differences in eye morphology.

Atlases are standardized reference images that are useful for tasks such as image registration and cross-sectional comparisons. For atlases to be representative of a population, it is important that they not be biased toward the morphology, contrast levels, or health conditions of any subject used in their creation. Given the variation across a population, it is challenging to generalize the population characteristics of both eye morphology and contrast intensity in a single anatomical reference template to define the conditional characteristics of the organ-specific regions (e.g., healthy or diseased). To enhance the generalization of eye organ contexts from different imaging protocols, we investigate the contextual variability in different tissue contrasts in MRI. Volumetric scans often have a lower resolution in the through-plane ($x$-$z$ or coronal plane and $y$-$z$ or sagittal plane) than that of the in-plane ($x$-$y$ or axial plane), where the $x$-axis is the left/right axis, the $y$-axis is the anterior/posterior axis, and the $z$-axis is the superior/inferior axis (Fig. 1). The low-resolution characteristics in the through-plane limit context for aligning the eye anatomies. Previous works have demonstrated the feasibility of leveraging deep learning super-resolution algorithms to restore the image quality.[9] To be useful for providing spatial context for low through-plane resolution MRI images of the eye, we need atlases that can appropriately visualize structures that are difficult to differentiate in low through-plane resolution, such as the optic nerve. Therefore, we desire to learn isotropic high-resolution information from images that contain only low-resolution information in the through-plane across several MRI tissue contrasts. Consequently, we explore two questions:

(1) Can we further apply a deep super-resolution algorithm to multiple MRI tissue contrasts?

(2) Can we leverage the super-resolution imaging to generate refined unbiased eye atlas templates?

In this work, we propose a coarse-to-fine framework to enhance the image resolution and leverage the restored details to generate a refined unbiased eye atlas specific to several tissue contrasts. We generate a separate atlas for each tissue contrast, so the atlases are not in spatial alignment. To represent the variability in eye morphology across a large population, we wish to incorporate information from as many subjects as possible. However, iterative deformable template generation algorithms are computationally expensive for more than a few subjects. To address this limitation, we choose a coarse-to-fine framework to create a coarse template from a small set of 25 subjects which we refine using a larger population of 75 subjects with a more computationally efficient deep learning-based deformable registration algorithm. The complete backbone consists of three steps: 1) applying a deep super-resolution network to enhance through-plane resolution quality; 2) generating an efficient coarse unbiased template from a small population of samples; and 3) refining the template by applying a deep probabilistic network for large population samples. The experimental results show that the application of the super-resolution network enhances the appearance of the eye organ. With the probabilistic refinement, our method achieves state-of-the-art registration performance when compared to deep learning registration baselines when there are sufficient subjects for refinement. Our contributions are summarized here:

(1) We propose a two-stage framework to enhance the through-plane resolution of imaging across different tissue contrasts and adapt the restored high-resolution context for eye atlas generation.

(2) We propose a coarse-to-fine registration strategy that combines both metric-based and deep learning-based registration to perform across large population samples.

(3) We evaluate our generated atlas with inverse eye organ label transfer from atlas space to moving subject space, demonstrating significant improvements in the Dice score across all tissue contrasts with sufficient subjects.

(4) All generated atlases as well as the corresponding four eye organ labels will be used through the Human BioMolecular Atlas Program (HuBMAP).[10]

The HuBMAP project highlights the need for standardized coordinate systems for navigating multiscale histological information in organs of the human body.[10] Here, the key contribution is a deep learning-based framework for generating eye atlases that provide this standardized coordinate system. We expand on previous work generating eye atlases for computed tomography to multi-contrast MRI acquired at low resolutions.[11] We contribute a pipeline for creating eye atlases using super-resolution and a coarse-to-fine framework for atlas generation. Here, we implement this method using the SMORE super-resolution algorithm along with the ANTs toolkit and VoxelMorph for deformable image registration.[9,12,13] Current eye atlases are generated using manual segmentation on rigid-aligned images.[14] A key contribution of the eye atlas proposed here is to provide a scaffold on which other information can be attached. For example, atlases allow for automated segmentation. We can register the template to a moving subject image and then apply this transformation to the atlas labels to label the moving subject image. Here, we introduce a pipeline for eye atlas generation and aim to establish a state-of-the-art method for eye atlases.

## 2 Related Works

*2.1 Atlas Generation*

Significant efforts have been dedicated to creating brain atlases, including across multiple modalities.[15] Researchers have created atlases with mouse brains to represent populational anatomy and variations.[16,17] Shi et al. developed an infant brain atlas, applying groupwise registration to avoid biasing the atlas to a single target.[18] There are multiple atlases that attempt to capture longitudinal information across infants of different ages,[19,20] with one using symmetric diffeomorphic registration to avoid bias.[21] While previous efforts primarily focused on creating healthy brain atlas templates, Rajashekar et al. proposed high-resolution normative atlases for visualizing population-wise representations of brain diseases, including brain lesion and stroke using fluid-attenuated inversion recovery (FLAIR) MRI and non-contrast CT modalities.[22] Abdominal studies have developed a multi-contrast kidney atlas, incorporating both contrast and morphological characteristics within kidney organs.[23,24] Researchers have extended kidney atlas templates to encompass substructure organs, such as the medulla, renal cortex, and pelvicalyceal systems in kidney regions using arterial phase CT.[25] However, limited research has addressed the creation of a standard reference atlas for the eye, which presents challenges due to its complex morphology and the influence of conditions that affect the eye shape, e.g., myopia and hyperopia.

*2.2 Medical Image Registration*

To accurately transfer the varied anatomical context from the moving subject to the atlas target, the image registration algorithm must be robust. One straightforward approach to enhancing registration performance is to adapt both affine and deformable transformations hierarchically with metric-guided optimization.[26–28] Furthermore, spatial optimization approaches attempt to regularize the deformation field to effectively align the anatomical context (e.g., discrete optimization,[29] b-spline deformation,[30] Demons,[31] and symmetric normalization[27]). However, the computational efficiency of these spatial transformations is limited.

Registration algorithms with deep neural networks aim to enhance both computational efficiency and robustness in an unsupervised setting. VoxelMorph is a foundational network that adapts a large deformation field to align the significant variation across anatomies.[28,32] Researchers have also adapted VoxelMorph to produce diffeomorphic deformations, i.e., deformations that are smooth and invertible.[32] To differentiate the two networks, we refer to the former as VoxelMorph-Original and the latter as VoxelMorph-Probabilistic. Zhao et al. crop the organ regions of interest (ROIs) and recursively register the anatomical context with VoxelMorph-Original,[33] while Yang et al. predict a bounding box to first localize the organ ROIs and perform registration.[34] Although deep learning-based approaches demonstrate their effectiveness to enhance the computational efficiency of registration algorithms, instability with the registration performance may arise due to substantial domain shifts with unseen data.[24]

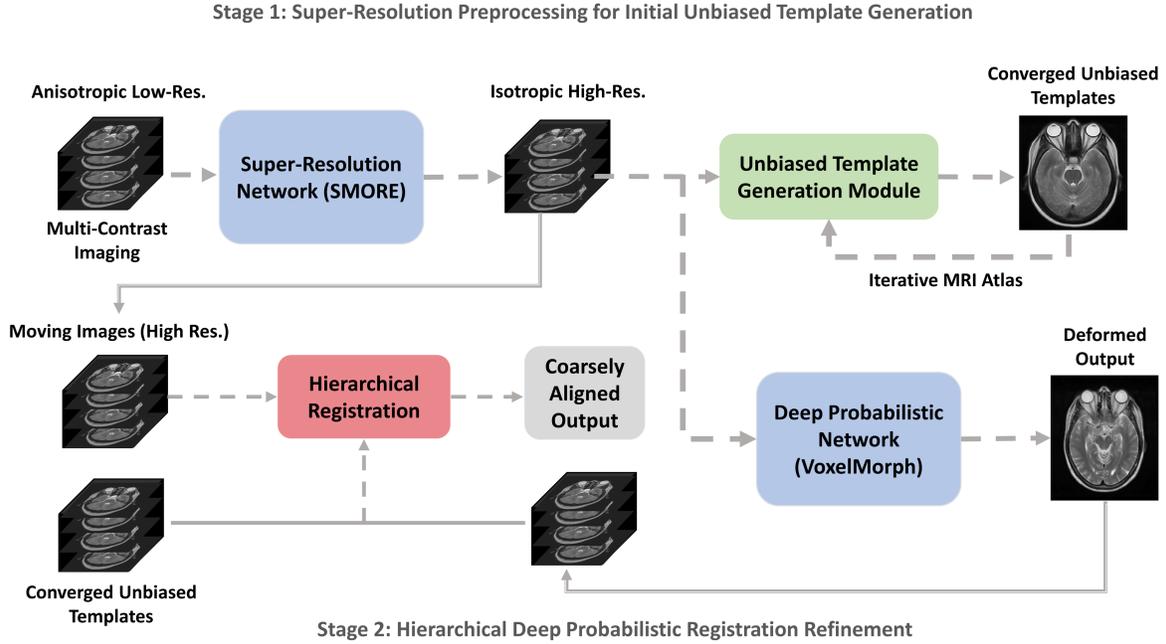

**Fig. 2** The complete pipeline for unbiased eye atlas generation consists of two stages: 1) performing a deep learning super-resolution algorithm to enhance image quality and distinguish organ appearances and 2) combining metric-based and deep learning-based registration through a hierarchical registration framework for refined anatomical transfer.

## 3 Methods

Our goal is to improve the through-plane resolution of different MRI tissue contrasts and leverage the distinct volumetric appearance in eye organs to generate tissue contrast-specific atlases across populations (Fig. 2). Our proposed framework can be divided into three sections: 1) super-resolution preprocessing, 2) coarse unbiased template generation, and 3) hierarchical deep probabilistic registration refinement.

*3.1 Super-Resolution Preprocessing*

We applied the synthetic multi-orientation resolution enhancement (SMORE) algorithm to generate super-resolution images.[9,35] We select SMORE as the super-resolution algorithm because it is self-supervised and does not require external training data. Other self-supervised super-resolution algorithms require orthogonal views of the same image across multiple contrasts[36] or train on a batch of images instead of each image independently.[37]

The input image for SMORE is an anisotropic volume, modeled with a spatial resolution of $l \times l \times h$, where $l$ and $h$ have units of mm and $h > l$. Here, the images have a high ratio between the in-plane resolution and through-plane resolution ($h/l > 6$). SMORE learns a correspondence between low-resolution (LR) and high-resolution (HR) image patches using only the in-plane slices as training data. The output of SMORE is an isotropic HR image with resolution $l \times l \times l$.

*3.2 Coarse Unbiased Template Generation*

Given the enriched context from the super-resolution algorithm in the prior step, we can now use the super-resolution images to create a generalized eye organ representation as a population-wise atlas template. Typically, we perform image registration to align and match the eye anatomy with imaging tools, e.g., ANTs and NiftyReg.[38,39] However, registration to a single target image with these tools is biased to a single fixed reference template.

To tackle this bias, we apply an unbiased template generation method that results in a coarse, generalized template despite the significant variance in eye morphology. Specifically, for each tissue contrast, we randomly sampled a small set of 25 subjects and generated an average mapping to coarsely align the skull region. The initial template is an average mapping of the 25 subjects, meaning it is unbiased to any of the subjects.[13,40] We performed hierarchical metric-based registration (consisting of rigid, affine, and then deformable registration) with ANTs to iteratively compute an average mapping in a separate spatial alignment for each tissue contrast. The computed average template in each epoch was the fixed template for the next epoch. We performed the same hierarchical procedure iteratively until the registration loss converged. We leveraged a small population sample to generate a coarse unbiased template due to the required time for loss convergence, which was 3 days for 20 samples and 3.5 weeks for 100 samples using an Intel Xeon W-2255 CPU. A previous study performed ANTs template generation on brain MRI using affine and deformable registration and found that two samples of 20 subjects each resulted in atlas templates with similar Jaccard scores for the whole brain and cortical regions, suggesting that this sample size is enough to average the variability across subjects for an initial template.[13] We hypothesize that the iterative-generated template can provide the representational anatomy of eye organs with minimal bias.

*3.3 Hierarchical Deep Probabilistic Registration Refinement*

We refined the template using the remaining randomly selected samples in addition to the 25 used for the coarse template generation. Our goal is to generalize the anatomical characteristics of eye organs across a large population. We used the VoxelMorph-Probabilistic model to refine the coarse atlas templates.[32] The deep probabilistic network predicts the deformation field modeled as a diffeomorphic transformation, meaning the transformation is smooth and invertible. Additionally, the model is unsupervised and does not require labels. In addition to the probabilistic model, we also compared with the non-probabilistic VoxelMorph-Original.[12] After refinement, the resulting atlases serve as reference images in separate spatial alignments for each tissue contrast. After forming the atlas template, we generate labels using majority voting.

Table 1 Overview of four multi-contrast MRI dataset samples

| Tissue Contrast | T1W Pre-Contrast | T1W Post-Contrast | T2W TSE | T2W FLAIR |
|---|---|---|---|---|
| Anatomical regions | Optic nerve, recti muscles, globe, orbital fat | | | |
| Sample Size | 44 | 100 | 100 | 100 |
| In-Plane Resolution (min-max, mm) | 0.430-0.938 | 0.375-0.938 | 0.391-0.898 | 0.393-0.898 |
| Slice Thickness (min-max, mm)* | 6.00 | 4.00-6.00 | 6.00 | 4.00-6.00 |

*This study used fully deidentified data. Information on the slice-selection profiles and use of slice gaps were removed in the deidentification process.

## 4   Experimental Setup

To evaluate our proposed unbiased atlas generation framework, we performed experiments to determine the quality of our super-resolution preprocessing and image registration pipeline. We tested our framework using inverse label transfer with four MRI tissue contrasts. We applied the inverse transformation using the deformation field of the atlas label and compared it to the original label for each subject. The choice of metrics and therefore performance for image analysis is highly application-specific.[41] Here, we choose to use Dice score to compare the inverse labels from atlas registered to subject with the original subject labels. We also calculated the Hausdorff distance both with and without super-resolution for each contrast to quantify the performance of distance-based metrics used to describe eye morphology.

*4.1 Datasets*

We retrieved de-identified volumetric scans in four different MRI tissue contrasts from 1842 patients from ImageVU, a medical image repository from Vanderbilt University Medical Center. We obtained approval from the Institutional Review Board (IRB 131461), and informed consent was waived due to the use of de-identified data. The tissue contrasts were T1-weighted pre-contrast, T1-weighted post-contrast, T2-weighted turbo-spin echo (TSE), and T2-weighted fluid attenuated inversion recovery (FLAIR). The ratio between the through-plane and in-plane resolution varied with a large level of ranges (Table 1). Across all four tissue contrasts studied here, the $x$-$y$ resolution varied from 0.457 mm to 0.635 mm, and the slice thickness varied from 1.23 mm to 7.00 mm. The large values for slice thickness limit our ability to distinguish spatial information. We randomly selected 100 subjects from each tissue contrast to both generate and evaluate the unbiased template, performing quality assurance to make sure the morphological conditions of the eyes are similar (e.g., healthy, no implant artifacts). For T1-weighted pre-contrast, there were only 44 total subjects. The subjects that we sampled for each imaging tissue contrast were different, resulting in different spatial alignments for each tissue contrast. All selected subject scans consisted of four organ ground truth labels: 1) optic nerve, 2) recti muscles, 3) globe and 4) orbital fat.

*4.2 Implementation Setup*

*4.2.1 Super-resolution preprocessing*

We apply the SMORE super-resolution algorithm to generated upsampled MRIs. After applying SMORE, we resampled the isotropic resolution to 0.8 mm × 0.8 mm × 0.8 mm using cubic interpolation. We further cropped and padded the MRI volumes to 256 × 256 × 224 voxels.

*4.2.2 Coarse unbiased template generation*

To generate the coarse unbiased template, we performed a conventional metric-based registration algorithm with the ANTs toolkit. We leveraged the multivariate template construction tool, which generates an average template that is not biased to a single subject. We applied both rigid and affine registration to align the anatomical locations of the head skull and eye organs, followed by SyN registration, which is a deformable registration algorithm with the similarity metric of cross-correlation. We chose four resolution levels (6, 4, 2, and 1) and iterated over each level for 100, 100, 70, and 20 iterations, respectively. We performed this registration process for six epochs and selected the generated template for each tissue contrast after the registration losses converged.

*4.2.3 Hierarchical registration refinement*

We used the remaining samples to refine the coarse template and generate a refined atlas template. As VoxelMorph-Original and VoxelMorph-Probabilistic assume the images only have non-linear spatial misalignment, we used the same hyperparameters in the template generation step to perform metric-based affine registration for the remaining samples as an initial registration alignment. Both the resolution and volumetric dimension of the MRI scans remained the same in the template generation stage (resolution: 0.8 mm × 0.8 mm × 0.8 mm, dimension: 256 × 256 × 224 voxels). We then trained the deep probabilistic framework available from VoxelMorph-Probabilistic and the non-probabilistic VoxelMorph-Original model for comparison. Due to hardware limitations, the batch size was 1. We used the Adam optimizer[42] with a learning rate of $10^{-4}$. Here, we chose to use the default hyperparameters for VoxelMorph and found the registration to be qualitatively satisfactory using a checkerboard visualization. A discussion of the impact of different hyperparameters can be found in studies by Balakrishnan et al. and Dalca et al.[12,32] For VoxelMorph, we used the original loss functions. For VoxelMorph-Original, we used normalized cross-correlation loss with a regularization term to encourage smooth displacement fields. For VoxelMorph-Probabilistic, we use KL divergence loss with normalized cross-correlation reconstruction loss. After the deep probabilistic refinement, we have a separate unbiased atlas for each tissue contrast.

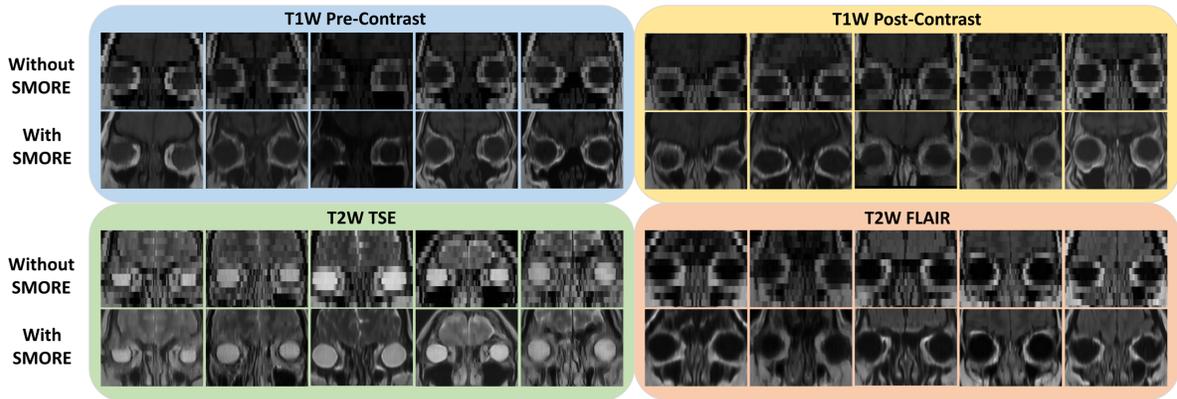

**Fig. 3** By applying SMORE (bottom rows), the anatomical context of the eye region is distinctly shown in the coronal view with a clear improvement in resolution across five unpaired patients in each tissue contrast compared to images without SMORE applied (top rows).

## 5  Results and Discussion

### 5.1  Qualitative Comparison with and without Super-Resolution Preprocessing

The super-resolution preprocessing enhanced the through-plane resolution images for each tissue contrast, with more distinctive appearances in eye organs (Fig. 3). The boundaries across tissues

Table 2 Quantitative evaluation of inverse transferred label for multiple eye organs across all patients

| Tissue Contrast | First Stage | Second Stage | Optic Nerve Dice Score | Recti Muscles Dice Score | Globe Dice Score | Orbital Fat Dice Score | Average Dice Score |
|---|---|---|---|---|---|---|---|
| T1W Pre-Contrast | ANTs | × | 0.828 ± 0.072 | 0.604 ± 0.188 | 0.737 ± 0.073 | **0.574 ± 0.153** | 0.686 ± 0.166 |
|  | ANTs | VoxelMorph-Original | **0.833 ± 0.071** | 0.601 ± 0.181 | 0.739 ± 0.073 | 0.570 ± 0.147 | **0.686 ± 0.165** |
|  | ANTs | VoxelMorph-Probabilistic | 0.828 ± 0.071 | **0.607 ± 0.184** | **0.740 ± 0.072** | 0.562 ± 0.145 | 0.684 ± 0.165 |
| T1W Post-Contrast | ANTs | × | 0.703 ± 0.190 | 0.498 ± 0.238 | 0.618 ± 0.159 | 0.364 ± 0.166 | 0.546 ± 0.229 |
|  | ANTs | VoxelMorph-Original | 0.772 ± 0.205 | **0.521 ± 0.212*** | 0.678 ± 0.16* | 0.442 ± 0.171* | 0.603 ± 0.228* |
|  | ANTs | VoxelMorph-Probabilistic | **0.773 ± 0.204** | 0.520 ± 0.217* | **0.680 ± 0.162*** | **0.443 ± 0.173*** | **0.604 ± 0.230*** |
| T2W TSE | ANTs | × | 0.733 ± 0.162 | 0.367 ± 0.229 | 0.672 ± 0.131 | 0.377 ± 0.168 | 0.538 ± 0.242 |
|  | ANTs | VoxelMorph-Original | **0.816 ± 0.160*** | 0.446 ± 0.214* | 0.741 ± 0.13* | 0.519 ± 0.165* | 0.631 ± 0.228* |
|  | ANTs | VoxelMorph-Probabilistic | 0.813 ± 0.159* | **0.451 ± 0.224*** | **0.743 ± 0.132*** | **0.520 ± 0.168*** | **0.632 ± 0.229*** |
| T2W FLAIR | ANTs | × | 0.742 ± 0.160 | 0.448 ± 0.243 | 0.666 ± 0.128 | 0.433 ± 0.143 | 0.572 ± 0.219 |
|  | ANTs | VoxelMorph-Original | 0.815 ± 0.175* | 0.579 ± 0.186* | 0.734 ± 0.139* | **0.584 ± 0.145*** | 0.678 ± 0.191* |
|  | ANTs | VoxelMorph-Probabilistic | **0.818 ± 0.175*** | **0.582 ± 0.184*** | **0.739 ± 0.14*** | 0.583 ± 0.139* | **0.681 ± 0.190*** |

*$p < 0.001$ using the Wilcoxon signed-rank test compared to ANTs alone. Note: bold values indicate highest mean Dice score for each label and contrast.

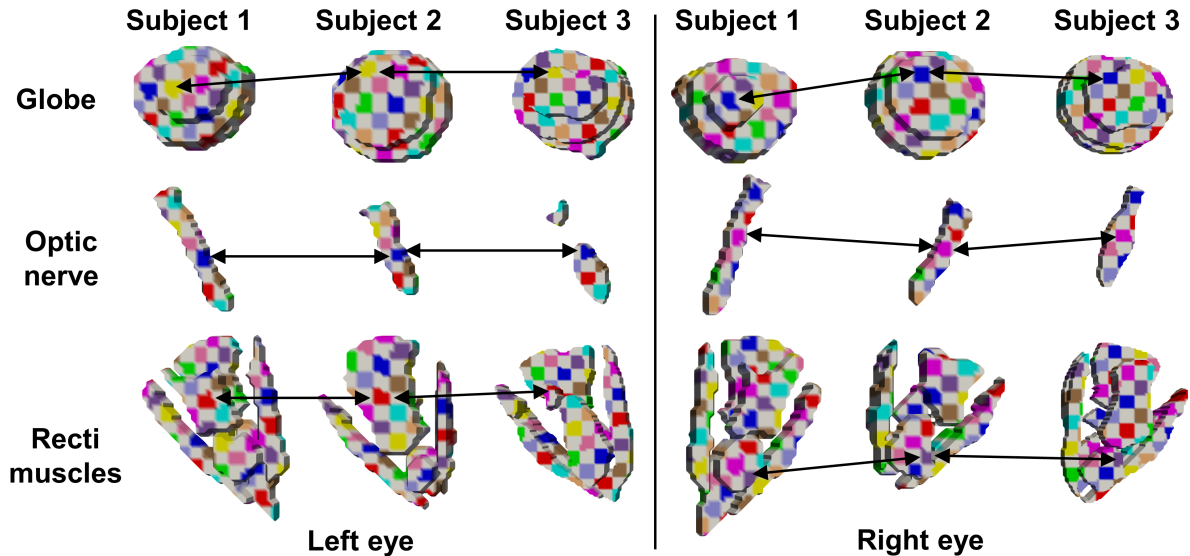

**Fig. 4** The atlas is generalizable across the variation in subjects, demonstrated by consistent registration for several subjects. The checkerboard shows the inverse deformation from atlas labels to moving subject labels for several subjects from the T2-weighted FLAIR tissue contrast. The arrows track a single square across subjects.

and anatomies are substantially clearer. This increase in image quality also demonstrates the distinctive variability of the eye organs across the population.

*5.2 Registration Comparisons across Multiple Contrast Images*

After we performed super-resolution preprocessing on all imaging cohorts, we performed hierarchical registration to align the anatomy from moving imaging samples to the unbiased atlas template. We applied ANTs as the first stage with a metric-based registration algorithm to create a baseline result across the four different tissue contrasts.

We performed the second stage registration using VoxelMorph-Original and VoxelMorph-Probabilistic (Table 2). We observed a statistically significant improvement in the Dice score across the four tissue contrasts using the Wilcoxon signed-rank test for all contrasts except T1-weighted pre-contrast, which had fewer subjects for refining the atlas. With the deep probabilistic model as the second stage, the label transfer performance significantly improved. The registration was consistent across the variable subjects (Fig. 4).

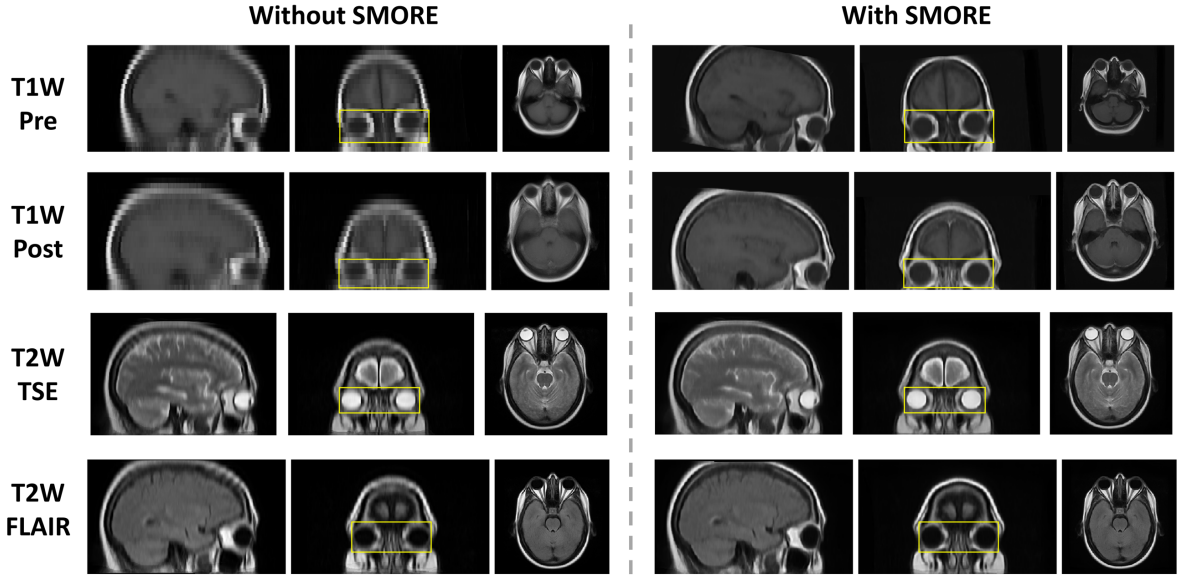

**Fig. 5** When using SMORE to generate an unbiased eye atlas, the anatomical context from eye organs to brain is refined, and tissues are clearly distinguishable compared to the unbiased eye atlas without using SMORE. The eye organ region (yellow bounding box) shows little deformation.

We observe that the unclear boundaries in the atlases brought by the low resolution in through-plane axis are minimized by applying SMORE (Fig. 5). The average Hausdorff distances for the inverse label transfer are approximately 6 mm, which is one voxel's thickness along the axial

Table 3 Quantitative evaluation of inverse label transfer with and without super-resolution using Hausdorff distance (HD)

| Tissue Contrast | Super-resolution? | Optic Nerve HD (mm) | Recti Muscles HD (mm) | Globe HD (mm) | Orbital Fat HD (mm) | Average HD (mm) |
|---|---|---|---|---|---|---|
| T1W Pre-Contrast | No | **4.70 ± 2.14** | 5.68 ± 1.87 | **3.69 ± 2.06** | 4.34 ± 1.97 | **4.60 ± 2.12** |
| | Yes | 4.85 ± 2.16 | **5.66 ± 1.58** | 4.71 ± 2.38 | **4.07 ± 1.61** | 4.82 ± 2.03 |
| T1W Post-Contrast | No | **5.99 ± 4.52** | **6.51 ± 5.23*** | **4.16 ± 6.25*** | **4.53 ± 5.39*** | **5.29 ± 5.45*** |
| | Yes | 6.61 ± 4.63 | 7.41 ± 5.50 | 6.20 ± 6.48 | 5.73 ± 5.73 | 6.48 ± 5.63 |
| T2W TSE | No | 7.39 ± 4.90 | **5.99 ± 1.49** | 5.29 ± 2.08 | **4.15 ± 1.76** | **5.70 ± 3.12** |
| | Yes | **7.09 ± 5.31** | 6.38 ± 3.52 | **4.98 ± 4.46*** | 4.67 ± 3.87 | 5.78 ± 4.44 |
| T2W FLAIR | No | 7.38 ± 11.53 | 8.09 ± 12.59 | 6.21 ± 14.46 | 6.85 ± 13.06 | 7.13 ± 12.92 |
| | Yes | **7.04 ± 11.76** | **7.35 ± 12.8** | **5.68 ± 14.67** | **5.33 ± 13.40*** | **6.35 ± 13.18*** |

*$p < 0.001$ using the Wilcoxon signed-rank test. Note: bold values indicate lowest Hausdorff distance for each label and contrast.

direction in the subject space. There is not a consistently significant difference in Hausdorff distance when performing super-resolution (Table 3). The mapping more clearly shows the anatomy of the eye organs and generalized population characteristics, with limited deformation in the eye organ region. A comparison of the inverse labels registered from atlas to moving subject space shows that the labels appear consistent with the original segmentation labels (Fig. 6).

*5.3 Discussion*

We presented a complete framework to adapt a large population of multi-contrast imaging for unbiased eye atlas generation. We integrated both metric-based and deep learning-based registration as a coarse-to-fine framework to refine the transfer process of eye organ anatomy across populations. By applying SMORE as the first step in the framework, the SMORE model learned the high-resolution context from the in-plane axial slice and applied the correspondence to restore the refined details for the through-plane coronal and sagittal slices. With the restored high-resolution details, the templates demonstrate a substantial qualitative enhancement in organ appearance and boundaries. However, there was not a consistently significant increase in Hausdorff distance using SMORE. This could be because the inverse label transfer involves registering a low-resolution subject image to the templates, limiting the spatial context available for registering the images regardless of the method used to generate the fixed template. With the rigid, affine, and deformable registration from ANTs, moving subject scans demonstrate coarse alignment with respect to the eye organs. The initial template is an average mapping that is not biased to a single subject, and each tissue contrast has a separate geometry. We further refined the intermediate registered output with a deep learning-based approach to generate a larger deformation field for anatomy alignment. Moreover, we integrated probabilistic neural networks to smooth the generated deformation field and to adapt diffeomorphism for registration, which enhanced the anatomical context transfer performance across all tissue contrasts with sufficient subjects.

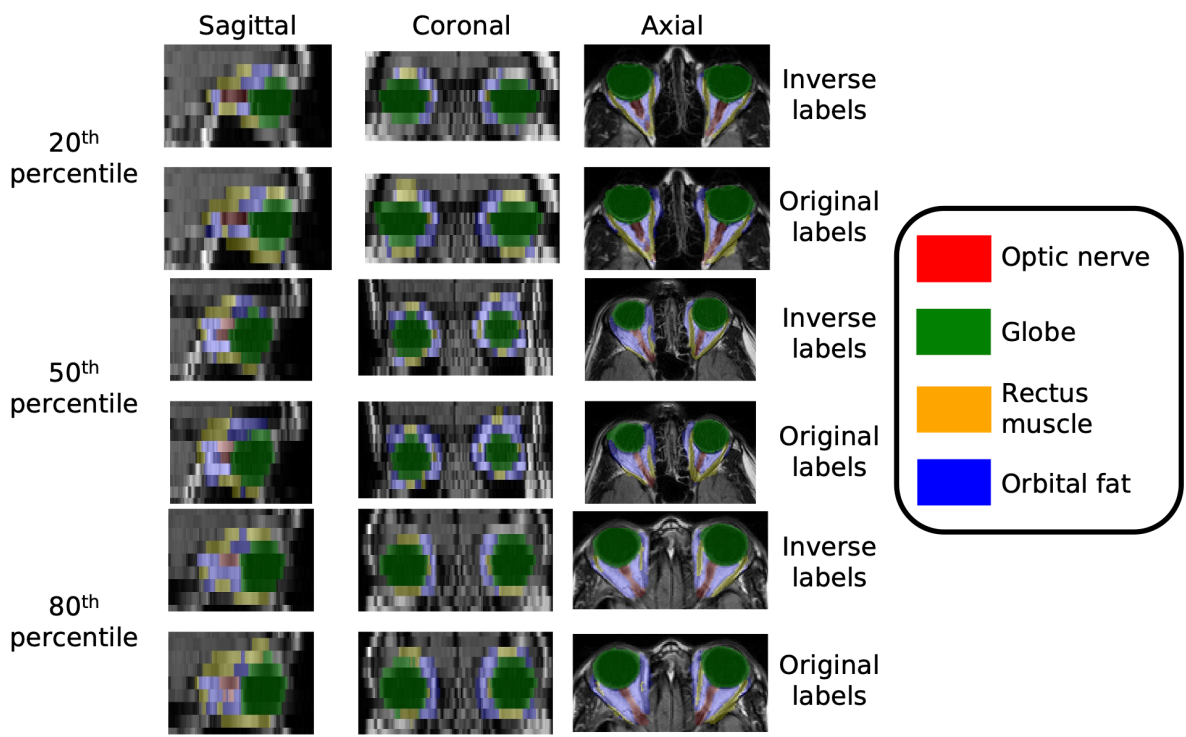

**Fig. 6** The inverse labels registered from the final atlas space to moving subject space appear qualitatively similar to the original segmentation labels. Here, we show examples of the labels on the moving subject MRI across the 20th, 50th, and 80th percentile of average Dice score across labels for the T2-weighted FLAIR tissue contrast.

Since the coarse template generation relies on an average mapping across 25 subjects, the atlases generated here are unbiased to a particular subject. This unbiased mapping addresses the limited information generalizable to a population from single subject atlases such as the Talairach-Tournoux human brain atlas.[43] There are several potential uses for these eye atlases. The main use is for the HuBMAP project, for localizing multi-scale information in the eye.[10] In medical research, they could be used to quantitatively measure eye shape across a variable population, similar to how brain atlases can allow for a standardized reference to quantify the volume of brain structures or size of small lesions. Atlases also allow for automatic labeling of structures of interest, providing confidence in images with poor quality.[44] Due to the application of the super-resolution algorithm, the eye atlases restore high resolution details that are not available in scans with a large slice thickness, meaning they provide a high-resolution reference for images with poor through-plane quality. The atlas generation pipeline also does not rely on any specific MRI tissue contrast, allowing for a consistent method for generating atlases across a broad range of tissue contrasts.

Although the generated unbiased templates for each tissue contrast demonstrate the distinctive appearance of the eye organs across population, multiple bottlenecks and limitations exist in the proposed framework. The first bottleneck is to generate a coarse unbiased template with ANTs. We only leveraged a small portion (25 subjects) of the imaging cohort to generate the initial average template. The main limitation of applying ANTs is the low computational efficiency, taking several days to generate a coarse template with only a small portion of samples, which can be a bottleneck without access to computing cluster resources. Therefore, an end-to-end approach to generate a coarse unbiased template is desirable. Another computational constraint is the hierarchical registration framework. Before applying deep learning-based registration algorithms like VoxelMorph-Original and VoxelMorph-Probabilistic, all imaging samples must be affine registered. However, limited studies have proposed adapting a deep learning network that can perform affine and deformable registration in parallel to avoid this sequential processing. Researchers have introduced multi-task networks combining affine and deformable registration to enhance the effectiveness and the computational efficiency of registration algorithms, but these networks have not shown substantial improvement over VoxelMorph-Original without the use of additional registration algorithms like Demons.[45] Another limitation of this framework is that the resulting atlases are not in registration, meaning we have a separate spatial geometry for each tissue contrast. Note that these computational limitations discussed here apply during the atlas construction. Since the atlases will be deployed offline outside of a clinical setting, computational concerns are secondary.

The framework presented here allows for the creation of a reference coordinate system for the eye. The eye atlases presented here provide a standardized coordinate system for histological information of the eye for use in the HuBMAP project.[10] The atlases allow for colocalization and navigation of multiscale information in the eye. Beyond this use, the eye atlases may also serve as a standardized spatial reference for the eye, serving as a means for exploring quantitative geometric measurements of eye morphology despite systematic differences within a population.

## 6 Conclusion

In summary, we introduced a framework to generate unbiased eye atlases across a large population using images with anisotropic voxels. We applied a deep learning super-resolution algorithm to learn the high-resolution characteristics from axial slices and applied this high-resolution correspondence to the coronal and sagittal slices. We adapted the restored high-resolution context to generate an unbiased eye atlas with a separate spatial geometry for each tissue contrast, using hierarchical registration with an average mapping to avoid biasing the atlas by registering to a single target. We integrated a deep probabilistic network to enhance the smoothness of the deformation field and increase registration performance with diffeomorphism. With sufficient subjects for refining the atlas, the generated average template from each tissue contrast illustrates the distinctive appearance of eye organs and generalizes across a large population cohort with significant improvement in anatomical label transfer performance compared to metric-based registration alone.

**Acknowledgments**

This research is supported by the NIH Common Fund U54 DK134302 and U54 EY032442 (Spraggins), NSF CAREER 1452485, NIH 2R01EB006136, NIH 1R01EB017230, NIH R01DK13557, NIH RO1NS09529, and NIH NIGMS T32GM007347 (Cho). This material is supported by the National Science Foundation Graduate Research Fellowship under Grant No. DGE-1746891 (SWR). ImageVU and RD are supported by the VICTR CTSA award (ULTR000445 from NCATS/NIH). This work was supported by Integrated Training in Engineering and Diabetes, grant number T32 DK101003. The Vanderbilt Institute for Clinical and Translational Research (VICTR) is funded by the National Center for Advancing Translational Sciences (NCATS) Clinical Translational Science Award (CTSA) Program, Award Number 5UL1TR002243-03. The content is solely the responsibility of the authors and does not necessarily represent the official views of the NIH. This work was conducted in part using the resources of the Advanced Computing Center for Research and Education at Vanderbilt University, Nashville, TN. We extend gratitude to NVIDIA for their support by means of the NVIDIA hardware grant. This work involved de-identified data obtained from human subjects. Approval was granted by the Institutional Review Board.

We use generative AI to create code segments based on task descriptions, as well as to debug, edit, and autocomplete code. Additionally, generative AI technologies have been employed to assist in structuring sentences and performing grammatical checks. The conceptualization, ideation, and all prompts provided to the AI originate entirely from the authors' creative and intellectual efforts. We take accountability for the review of all content generated by AI in this work.

*Code and Data Availability*

The code for the tools used here are available online: ANTs at https://stnava.github.io/ANTs/, VoxelMorph at https://github.com/voxelmorph/voxelmorph, and SMORE at https://gitlab.com/iacl/smore. The atlases will be available through the HuBMAP project.[10]

**References**

1. B. Tan et al., "Ultrawide field, distortion-corrected ocular shape estimation with MHz optical coherence tomography (OCT)," Biomed Opt Express **12**(9), 5770 (2021) [doi:10.1364/BOE.428430].

2. L. S. Lim et al., "MRI of posterior eye shape and its associations with myopia and ethnicity," British Journal of Ophthalmology, bjophthalmol-2019-315020 (2019) [doi:10.1136/bjophthalmol-2019-315020].

3. R. Aseem et al., "Positional Variation of the Infraorbital Foramen in Caucasians and Black Africans from Britain: Surgical Relevance and Comparison to the Existing Literature," Journal of Craniofacial Surgery **32**(3), 1162–1165 (2021) [doi:10.1097/SCS.0000000000007014].

4. D. Dean et al., "Average African American Three-Dimensional Computed Tomography Skull Images," Journal of Craniofacial Surgery **9**(4), 348–358 (1998) [doi:10.1097/00001665-199807000-00011].

5. D. H. Kim, J.-S. Jun, and R. Kim, "Ultrasonographic measurement of the optic nerve sheath diameter and its association with eyeball transverse diameter in 585 healthy volunteers," Sci Rep **7**(1), 15906 (2017) [doi:10.1038/s41598-017-16173-z].

6. M. Vaiman, R. Abuita, and I. Bekerman, "Optic nerve sheath diameters in healthy adults measured by computer tomography," Int J Ophthalmol **8**(6), 1240–1244 (2015).
7. D. A. Atchison et al., "Eye Shape in Emmetropia and Myopia," Investigative Opthalmology & Visual Science **45**(10), 3380 (2004) [doi:10.1167/iovs.04-0292].
8. I. Bekerman, P. Gottlieb, and M. Vaiman, "Variations in Eyeball Diameters of the Healthy Adults," J Ophthalmol **2014**, 1–5 (2014) [doi:10.1155/2014/503645].
9. C. Zhao et al., "SMORE: A Self-Supervised Anti-Aliasing and Super-Resolution Algorithm for MRI Using Deep Learning," IEEE Trans Med Imaging **40**(3), 805–817 (2021) [doi:10.1109/TMI.2020.3037187].
10. S. Jain et al., "Advances and prospects for the Human BioMolecular Atlas Program (HuBMAP)," Nat Cell Biol (2023) [doi:10.1038/s41556-023-01194-w].
11. H. H. Lee et al., "Unsupervised registration refinement for generating unbiased eye atlas," in Medical Imaging 2023: Image Processing, I. Išgum and O. Colliot, Eds., p. 77, SPIE (2023) [doi:10.1117/12.2653753].
12. G. Balakrishnan et al., "VoxelMorph: A Learning Framework for Deformable Medical Image Registration," IEEE Trans Med Imaging **38**(8), 1788–1800 (2019) [doi:10.1109/TMI.2019.2897538].
13. B. B. Avants et al., "A Reproducible Evaluation of ANTs Similarity Metric Performance in Brain Image Registration," Neuroimage **54**(3), 2033, NIH Public Access (2011) [doi:10.1016/J.NEUROIMAGE.2010.09.025].
14. D. B. P. Eekers et al., "Update of the EPTN atlas for CT- and MR-based contouring in Neuro-Oncology," Radiotherapy and Oncology **160**, 259–265, Elsevier (2021) [doi:10.1016/J.RADONC.2021.05.013].
15. P. Lorenzen et al., "Multi-modal image set registration and atlas formation," Med Image Anal **10**(3), 440–451, Elsevier (2006) [doi:10.1016/J.MEDIA.2005.03.002].
16. N. Kovačević et al., "A Three-dimensional MRI Atlas of the Mouse Brain with Estimates of the Average and Variability," Cerebral Cortex **15**(5), 639–645 (2005) [doi:10.1093/cercor/bhh165].

17. Q. Wang et al., "The Allen Mouse Brain Common Coordinate Framework: A 3D Reference Atlas," Cell **181**(4), 936-953.e20, Cell Press (2020) [doi:10.1016/J.CELL.2020.04.007].

18. F. Shi et al., "Infant Brain Atlases from Neonates to 1- and 2-Year-Olds," PLoS One **6**(4), e18746 (2011) [doi:10.1371/journal.pone.0018746].

19. Y. Zhang et al., "Consistent Spatial-Temporal Longitudinal Atlas Construction for Developing Infant Brains," IEEE Trans Med Imaging **35**(12), 2568–2577 (2016) [doi:10.1109/TMI.2016.2587628].

20. M. Kuklisova-Murgasova et al., "A dynamic 4D probabilistic atlas of the developing brain," Neuroimage **54**(4), 2750–2763, Academic Press (2011) [doi:10.1016/J.NEUROIMAGE.2010.10.019].

21. A. Gholipour et al., "A normative spatiotemporal MRI atlas of the fetal brain for automatic segmentation and analysis of early brain growth," Sci Rep **7**(1), 476 (2017) [doi:10.1038/s41598-017-00525-w].

22. D. Rajashekar et al., "High-resolution T2-FLAIR and non-contrast CT brain atlas of the elderly," Sci Data **7**(1), 56 (2020) [doi:10.1038/s41597-020-0379-9].

23. H. H. Lee et al., "Construction of a multi-phase contrast computed tomography kidney atlas," in Medical Imaging 2021: Image Processing, B. A. Landman and I. Išgum, Eds., p. 61, SPIE (2021) [doi:10.1117/12.2580561].

24. H. H. Lee et al., "Multi-contrast computed tomography healthy kidney atlas," Comput Biol Med **146**, 105555, Pergamon (2022) [doi:10.1016/J.COMPBIOMED.2022.105555].

25. H. H. Lee et al., "Supervised deep generation of high-resolution arterial phase computed tomography kidney substructure atlas," in Medical Imaging 2022: Image Processing, I. Išgum and O. Colliot, Eds., p. 97, SPIE (2022) [doi:10.1117/12.2608290].

26. J. Ashburner, "A fast diffeomorphic image registration algorithm," Neuroimage **38**(1), 95–113, Academic Press (2007) [doi:10.1016/J.NEUROIMAGE.2007.07.007].

27. B. B. Avants et al., "Symmetric diffeomorphic image registration with cross-correlation: Evaluating automated labeling of elderly and neurodegenerative brain," Med Image Anal **12**(1), 26–41, Elsevier (2008) [doi:10.1016/J.MEDIA.2007.06.004].

28. G. Balakrishnan et al., "An Unsupervised Learning Model for Deformable Medical Image Registration," in 2018 IEEE/CVF Conference on Computer Vision and Pattern Recognition, pp. 9252–9260, IEEE (2018) [doi:10.1109/CVPR.2018.00964].

29. A. V. Dalca et al., "Patch-Based Discrete Registration of Clinical Brain Images," 60–67 (2016) [doi:10.1007/978-3-319-47118-1_8].

30. D. Rueckert et al., "Nonrigid registration using free-form deformations: application to breast MR images," IEEE Trans Med Imaging **18**(8), 712–721 (1999) [doi:10.1109/42.796284].

31. T. Vercauteren et al., "Diffeomorphic demons: Efficient non-parametric image registration," Neuroimage **45**(1), S61–S72, Academic Press (2009) [doi:10.1016/J.NEUROIMAGE.2008.10.040].

32. A. V. Dalca et al., "Unsupervised learning of probabilistic diffeomorphic registration for images and surfaces," Med Image Anal **57**, 226–236, Elsevier (2019) [doi:10.1016/J.MEDIA.2019.07.006].

33. S. Zhao et al., "Recursive Cascaded Networks for Unsupervised Medical Image Registration," in 2019 IEEE/CVF International Conference on Computer Vision (ICCV), pp. 10599–10609, IEEE (2019) [doi:10.1109/ICCV.2019.01070].

34. S. di Yang et al., "Target organ non-rigid registration on abdominal CT images via deep-learning based detection," Biomed Signal Process Control **70**, 102976, Elsevier (2021) [doi:10.1016/J.BSPC.2021.102976].

35. S. W. Remedios et al., "Self-Supervised Super-Resolution for Anisotropic MR Images with and Without Slice Gap," 118–128, Springer, Cham (2023) [doi:10.1007/978-3-031-44689-4_12].

36. J. McGinnis et al., "Single-subject Multi-contrast MRI Super-resolution via Implicit Neural Representations," pp. 173–183 (2023) [doi:10.1007/978-3-031-43993-3_17].

37. H. Zhang et al., "Self-supervised arbitrary scale super-resolution framework for anisotropic MRI" (2023).

38. B. B. Avants et al., "The Insight ToolKit image registration framework," Front Neuroinform **8** (2014) [doi:10.3389/fninf.2014.00044].

39. M. Modat et al., "Global image registration using a symmetric block-matching approach," Journal of Medical Imaging **1**(2), 024003 (2014) [doi:10.1117/1.JMI.1.2.024003].

40. B. B. Avants et al., "The Optimal Template Effect in Hippocampus Studies of Diseased Populations," Neuroimage **49**(3), 2457, NIH Public Access (2010) [doi:10.1016/J.NEUROIMAGE.2009.09.062].

41. L. Maier-Hein et al., "Metrics reloaded: recommendations for image analysis validation," Nat Methods **21**(2), 195–212 (2024) [doi:10.1038/s41592-023-02151-z].

42. D. P. Kingma and J. Ba, "Adam: A Method for Stochastic Optimization," in 3rd International Conference for Learning Representations (2014).

43. D. A. Dickie et al., "Whole Brain Magnetic Resonance Image Atlases: A Systematic Review of Existing Atlases and Caveats for Use in Population Imaging," Front Neuroinform **11** (2017) [doi:10.3389/fninf.2017.00001].

44. W. L. Nowinski, "Usefulness of brain atlases in neuroradiology: Current status and future potential," Neuroradiol J **29**(4), 260, SAGE Publications (2016) [doi:10.1177/1971400916648338].

45. X. Gao et al., "DeepASDM: a Deep Learning Framework for Affine and Deformable Image Registration Incorporating a Statistical Deformation Model," in 2021 IEEE EMBS International Conference on Biomedical and Health Informatics (BHI), pp. 1–4, IEEE (2021) [doi:10.1109/BHI50953.2021.9508553].

**Ho Hin Lee** earned his PhD in computer science from Vanderbilt University in 2023, as well as an MS degree in biomedical engineering from Columbia University in 2019 and a BE in biomedical engineering from the Chinese University of Hong Kong in 2013. His interests include machine learning, medical image analysis, and biomedical representation learning.

**Adam Saunders** is a PhD student in electrical and computer engineering at Vanderbilt University. He earned a BEE degree from University of Dayton in 2023. His current research interests include deep learning applications in medical imaging and quantitative imaging methods for MRI.

Biographies and photographs for the other authors are not available.